\documentclass[twocolumn]{aastex702}
\usepackage{placeins}
\usepackage{amsmath}

\newcommand{\Cthree}{\ensuremath{\mathcal{C}_{3.3}}}
\newcommand{\Xthree}{\ensuremath{\mathcal{X}_{3.3}}}
\newcommand{\Ithreewin}{\ensuremath{I_{3.3,\rm win}}}
\newcommand{\Ieight}{\ensuremath{I_{8,\rm dust}}}
\newcommand{\Cbralpha}{\ensuremath{\mathcal{C}_{\rm Br\alpha}}}
\newcommand{\Xbralpha}{\ensuremath{\mathcal{X}_{\rm Br\alpha}}}
\newcommand{\Deltathree}{\ensuremath{\Delta_{3.3}}}
\newcommand{\Deltaeight}{\ensuremath{\Delta_8}}
\newcommand{\sigmad}{\ensuremath{\Sigma_{\rm d}}}
\newcommand{\bralpha}{\ensuremath{{\rm Br}\alpha}}
\newcommand{\halpha}{\ensuremath{{\rm H}\alpha}}
\newcommand{\tir}{\ensuremath{\mathrm{TIR}}}
\newcommand{\qpah}{\ensuremath{q_{\rm PAH}}}
\newcommand{\um}{\ensuremath{\mu{\rm m}}}

\shorttitle{SPHEREx 3.3 Micron Aromatic Emission in the Magellanic Clouds}
\shortauthors{Li}

\begin{document}

\title{SPHEREx 3.3 Micron Aromatic Emission in the Magellanic Clouds: Separating Dust Column, Excitation, and Environmental Suppression}

\author[0000-0002-8711-8970]{Cheng Li}
\affiliation{Department of Astronomy, Tsinghua University, Beijing 100084, China}
\email{cli2015@tsinghua.edu.cn}
\correspondingauthor{Cheng Li}
\email{cli2015@tsinghua.edu.cn}

\begin{abstract}
Bright emission in the 3.3 \um\ unidentified infrared band (UIB), often described as an aromatic infrared band, can trace dust column, radiative excitation, or the survival and emissivity of its carriers.  We combine SPHEREx maps of the Magellanic Clouds with HERITAGE dust columns, a TIR heating proxy, Br$\alpha$, H$\alpha$, H{\sc ii}-region catalogs, and SAGE 8 \um\ imaging.  We measure continuum-subtracted 3.3 \um\ feature excess, \Xthree, and Br$\alpha$ excess, \Xbralpha, in matched cells.  The Clouds are analyzed symmetrically but sample different regimes: the LMC supplies most high-\Xbralpha\ active-region dynamic range, whereas the SMC probes a lower-metallicity, lower-surface-brightness regime.  In each Cloud, a quiescent diffuse baseline depending on dust surface density, \sigmad, and TIR/$\sigmad$ predicts \Xthree\ with $R^2=0.628$ in the LMC and 0.288 in the SMC.  Relative to this expectation, LMC H{\sc ii} interiors are under-luminous by $-0.223^{+0.035}_{-0.040}$ dex; local-background measurements give $-0.153^{+0.014}_{-0.043}$ dex despite a total-\Xthree\ enhancement of $+0.749$ dex.  SMC H{\sc ii}-region residuals are consistent with zero over its lower-\Xbralpha\ regime.  A matched comparison with stellar-subtracted 8 \um\ dust intensity, \Ieight, shows that column and heating predict \Ieight\ more tightly than \Xthree, and that the high-\Xbralpha\ LMC deficit is stronger for \Xthree\ than for \Ieight.  The deficit is therefore selective to the isolated 3.3 \um\ feature, not uniform in all aromatic-band-dominated emission.  This result suggests that the small aromatic/aliphatic carbonaceous carriers traced by the isolated 3.3 \um\ feature are more susceptible to processing in active regions than the larger or differently emitting carbonaceous-grain population dominating the 8 \um\ aromatic complexes.
\end{abstract}


\section{Introduction}

The 3.3 \um\ emission feature belongs to the family of unidentified infrared bands (UIBs), now usually described as aromatic infrared bands, first recognized in dusty Galactic nebulae and then in galaxies \citep{GillettForrestMerrill1973,RussellSoiferMerrill1977,WillnerPuetterRussellSoifer1979}.  Polycyclic aromatic hydrocarbons (PAHs) provide the leading interpretation: in this picture, the 3.3 \um\ band traces aromatic C--H stretching, while the 6--12 \um\ complexes trace other aromatic vibrational modes \citep{LegerPuget1984,Allamandola1985,Allamandola1989,Tielens2008,Peeters2021}.  The astronomical carrier is generally treated as an ensemble rather than a single identified molecule, and the 3.2--3.6 \um\ complex has also been discussed in terms of a broader carbonaceous-carrier family: PAH clusters and very small grains, chemically modified or substituted PAHs, hydrogenated amorphous carbon and quenched carbonaceous composite materials, coal/kerogen-like macromolecular organics, and mixed aromatic--aliphatic organic nanoparticles \citep{Knacke1977,DuleyWilliams1981,Sakata1990,ScottDuleyJahani1997,Papoular1989,Papoular2001,Rapacioli2005,Pilleri2012,KwokZhang2011,KwokZhang2013,Yang2013,Yang2016,Yang2017,Sadjadi2017,Kwok2022}.

Aromatic-band emission is widely used to trace the dusty interstellar medium (ISM), star formation, and dust processing in galaxies \citep{Tielens2008,Smith2007}.  Its observed strength depends on several coupled quantities: the abundance of aromatic carbonaceous grains, the amount of dust along the line of sight, the local radiation field, grain charge and size distribution, and destruction or dilution in harsh environments \citep[e.g.,][]{DraineLi2007,Draine2007,Gordon2008}.  This complexity is especially important in low-metallicity systems, where aromatic/PAH emission is known to be weak and environmentally structured \citep{Roche1991,Engelbracht2005,Madden2006,Galliano2008,Sandstrom2010,Sandstrom2012,Whitcomb2024,Whitcomb2026}.

The Magellanic Clouds are the nearest laboratories where these effects can be separated on resolved physical scales.  SAGE/SAGE-SMC and HERITAGE provide infrared and dust constraints across both Clouds \citep{Meixner2006,Gordon2011,Meixner2013,Gordon2014}, and previous work shows that PAH mass fractions, 8 \um\ emission, and infrared colors vary strongly around active regions and H{\sc ii} complexes \citep[e.g.,][]{Sandstrom2010,Sandstrom2012,Oey2017,Chastenet2019}.  ISO and AKARI spectroscopy further found LMC aromatic-band-ratio variations, including changes in the 3.3 \um\ feature, interpreted in terms of excitation and small-carrier processing \citep{Vermeij2002,Mori2012}.  Broadband mid-infrared diagnostics, however, mix aromatic features, stellar continuum, very small grains, and warm dust continuum \citep{Helou2004,Calzetti2007,Bendo2008,Boquien2010}.  A remaining observational problem is whether the isolated 3.3 \um\ feature responds like broader aromatic-band-dominated mid-infrared emission once dust column and heating are controlled.

SPHEREx provides all-sky low-resolution spectra from the near- to mid-infrared \citep{Dore2014,Dore2016,Dore2018,Korngut2018,Crill2020,Bock2026,Korngut2026,Hui2026}, enabling direct maps of the 3.3 \um\ aromatic feature and Br$\alpha$ at 4.05 \um.  The 3.3 \um\ feature is now a key aromatic/UIB diagnostic for JWST and SPHEREx studies of nearby and distant galaxies \citep{Kim2012,Lai2020,Rigopoulou2021,Spilker2023,Sandstrom2023PAH3,Koziol2026,Zhang2025,Hora2026}, while JWST and Galactic-plane SPHEREx results already show strong aromatic-feature/recombination-line structure around ionized gas \citep{Chastenet2023PAHFraction,Egorov2023,Pedrini2024,Gregg2024,Murgia2026}.  The remaining test is whether this separation can be quantified across entire external, low-metallicity galaxies after controlling for dust column and heating.  The Magellanic Clouds provide this joint test, with the LMC and SMC sampling different dust surface brightness, ionized-gas activity, and star-forming structure.

This paper addresses that problem by asking what controls the resolved 3.3 \um\ feature in the Magellanic Clouds: aromatic-carrier column, radiative excitation, ionized-gas activity, or carrier survival and emissivity.  We combine SPHEREx 3.3 \um\ and Br$\alpha$ maps with HERITAGE dust columns, a far-infrared TIR/$\sigmad$ heating proxy, and SAGE 8 \um\ aromatic-band-dominated emission.  The directly isolated 3.3 \um\ feature is compared with stellar-subtracted SAGE 8 \um\ emission, dominated by the 7.7 and 8.6 \um\ aromatic complexes \citep{Helou2004,Smith2007,Bendo2008}.  This work separates dust column, heating, and residual environmental behavior relative to a quiescent diffuse baseline, then tests whether the residual is common to aromatic-band-dominated emission or stronger in the small-carrier-sensitive 3.3 \um\ feature.

The remainder of this paper is organized as follows.  Section~\ref{sec:data} describes the SPHEREx, SAGE, HERITAGE, far-infrared, H$\alpha$, and H{\sc ii}-region data products.  Section~\ref{sec:methods} defines the cell samples, diffuse-baseline decompositions for \Xthree\ and \Ieight, and local-background measurements.  Section~\ref{sec:results} presents the full-cloud maps, the dust-column/heating/Br$\alpha$ relations, and the environmental residual comparison between the 3.3 and 8 \um\ aromatic bands.  Section~\ref{sec:discussion} discusses the implications for aromatic-carrier excitation and processing, and Section~\ref{sec:summary} summarizes the main conclusions.  Appendix~\ref{app:product_validation} presents the product-validation and robustness tests.  When physical scales are quoted, we adopt distances of 49.59 kpc for the LMC \citep{Pietrzynski2019} and 62.44 kpc for the SMC \citep{Graczyk2020}, so that 1 arcmin corresponds to 14.4 pc and 18.2 pc, respectively.  Logarithms are base 10.

\section{Data}
\label{sec:data}

\subsection{SPHEREx 3.3 Micron Aromatic Feature and \texorpdfstring{Br$\alpha$}{Br-alpha} Maps}

We use public calibrated SPHEREx spectral-image products processed by the SPHEREx Science Data Center \citep{Akeson2025}.  The raw L2 cutouts used in this paper, covering the LMC and SMC fields, were downloaded from the public archive as of 2026 July 25.  D4 spectra measure the 3.3 \um\ aromatic feature and D5 spectra measure Br$\alpha$ \citep{Bock2026,Hui2026}.  The fiducial feature windows are 3.18--3.40 \um\ for the 3.3 \um\ feature and 4.00--4.10 \um\ for Br$\alpha$, with local continua fit from bracketing sidebands.  For each sky tile we form continuum, excess, uncertainty, and S/N maps, mosaic the LMC and SMC separately, and sample the mosaics onto common matched cells.

For this work we developed a SPHEREx map-making pipeline, SPIRAL (SPHEREx Pipeline for Infrared Recombination and Aromatic-Line mapping).\footnote{\url{https://github.com/cli-tsinghua/SPIRAL}}  This pipeline follows the spectral map-making procedure of \citet{Cukierman2026}, adapted to the LMC/SMC fields and the two feature windows.  For each calibrated cutout it subtracts the zodiacal foreground, masks flagged or invalid pixels, bins valid pixels onto 2 and 4 arcmin cloud grids, fits local linear continua, and records continuum, integrated excess, uncertainty, S/N, equivalent width, and provenance.

Appendix~\ref{app:product_validation} summarizes validation against IRSA SPHEREx mosaics, point-source photometry, closure tests, injection/recovery tests, negative-control windows, split-sample repeatability, and pipeline variants.  The products have consistent morphology, astrometry, and mean flux scale at the adopted cell scale; low-level coherent residual texture is treated as a systematic floor.  Negative continuum-subtracted excesses and S/N values are interpreted as noise and continuum-placement residuals, not physical absorption.

Throughout this paper, $\mathcal{C}$ denotes the fitted local continuum surface brightness and $\mathcal{X}$ denotes the continuum-subtracted integrated feature excess.  Thus \Cthree\ and \Cbralpha\ are the continuum maps near the two features, while \Xthree\ and \Xbralpha\ are the 3.3 \um\ feature and Br$\alpha$ excess maps used in the quantitative measurements.

Smooth ionized-gas continuum tied to Br$\alpha$ is too small to affect the measured 3.3 \um\ feature excess.  We scale an optically thin hydrogen free-free spectrum to the Case-B Br$\alpha$ emissivity \citep{HummerStorey1987} and pass it through the same 3.18--3.40 \um\ feature window and 2.95--3.16 and 3.60--3.82 \um\ continuum sidebands.  At $T_e=10^4$ K, the full free-free continuum integrated across the feature window is $0.201\Xbralpha$ before continuum subtraction, but the local linear continuum fit removes 99.7\% of this smooth component, leaving a curvature residual of only $6.0\times10^{-4}\Xbralpha$.  Across $T_e=5000$--20,000 K, the largest residual coefficient is $8.0\times10^{-4}\Xbralpha$.  For H{\sc ii}-interior cells this conservative upper bound gives median biases of $4.2\times10^{-4}\Xthree$ in the LMC and $7.1\times10^{-4}\Xthree$ in the SMC, with 95th percentiles of 0.23\% and 0.49\%, respectively.  No free-free continuum correction is applied.

We use \Xbralpha\ as an ionized-gas and massive-star-feedback tracer.  Because a positive \Xthree--\Xbralpha\ correlation can reflect excitation or common star-forming structure, we first model \Xthree\ with dust column and heating, then test whether the residuals are organized by ionized-gas environment.

\subsection{Dust, Heating, and Ionized-Gas Tracers}

We use SAGE/SAGE-SMC IRAC 3.6 and 8.0 \um\ maps to construct the stellar-continuum-subtracted dust intensity $\Ieight=I_8-0.232I_{3.6}$ \citep{Helou2004}.  \Ieight\ is a broadband aromatic-band-dominated control, not a PAH mass fraction, and tests whether \Xthree\ behaves like the stronger 7.7 and 8.6 \um\ aromatic complexes.  We use SAGE/SAGE-SMC MIPS 24, 70, and 160 \um\ maps only to construct a TIR-like heating proxy following \citet{DaleHelou2002}.

Dust-column information comes from HERITAGE dust-model products \citep{Meixner2013,Gordon2014,Chastenet2019}.  Our baseline model uses $\sigmad$ as the column term and $\tir/\sigmad$ as a proxy for radiative heating per unit dust mass, in the same spirit as dust-emission models that separate dust mass, PAH fraction, and starlight intensity \citep{DraineLi2007,Draine2007}.  H{\sc ii} region centers and radii define region interiors and local-background annuli \citep{Lopez2014}, while SHASSA H$\alpha$ maps help identify ionized material outside cataloged H{\sc ii} interiors \citep{Gaustad2001}.  The \halpha\ and \bralpha\ tracers are complementary: \halpha\ gives sensitive, high-coverage morphology but is affected by attenuation and scattering, whereas \bralpha\ is less extinguished and measured in the same SPHEREx data stream as the 3.3 \um\ feature.  We use \halpha\ mainly for environment definition and \Xbralpha\ as a co-spatial infrared tracer of ionized gas and massive-star feedback \citep{HummerStorey1987,KennicuttEvans2012}.

\begin{figure*}[!t]
\centering
\includegraphics[width=0.98\textwidth]{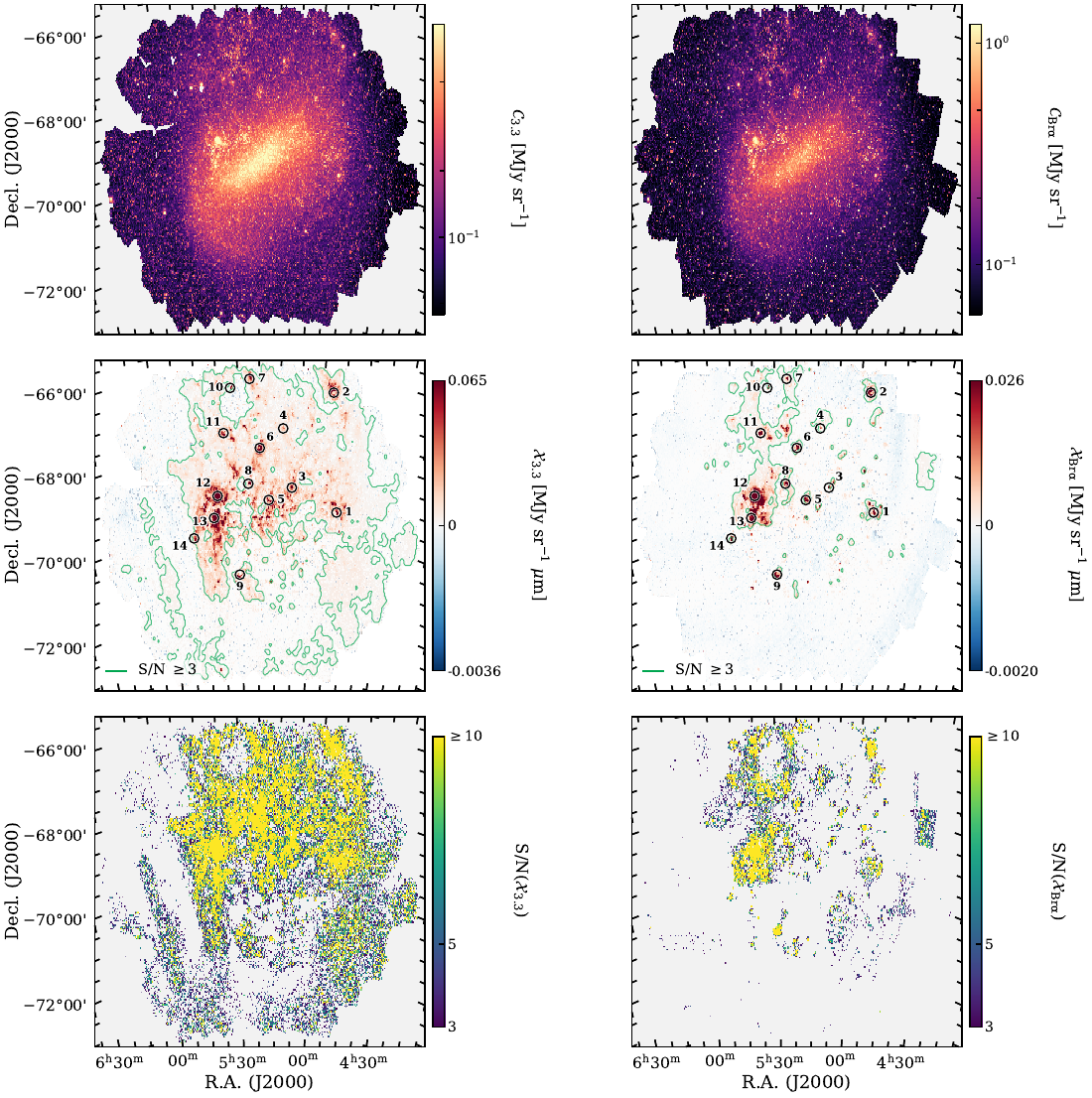}
\caption{Full-cloud SPHEREx maps of the LMC.  Columns show the 3.3 \um\ aromatic feature and Br$\alpha$; rows show the fitted continuum $\mathcal{C}$, continuum-subtracted excess $\mathcal{X}$, and feature S/N.  The $\mathcal{X}$ panels retain the signed excess map; green contours mark coherent S/N$\geq3$ detections, and open circles with cloud-specific numeric IDs mark the fiducial resolved H{\sc ii} regions listed in Table~\ref{tab:hii_regions_individual}.  The S/N panels show positive detections only, clipped at S/N=10.}
\label{fig:lmc_full_spherex}
\end{figure*}

\begin{figure*}[t]
\centering
\includegraphics[width=0.98\textwidth]{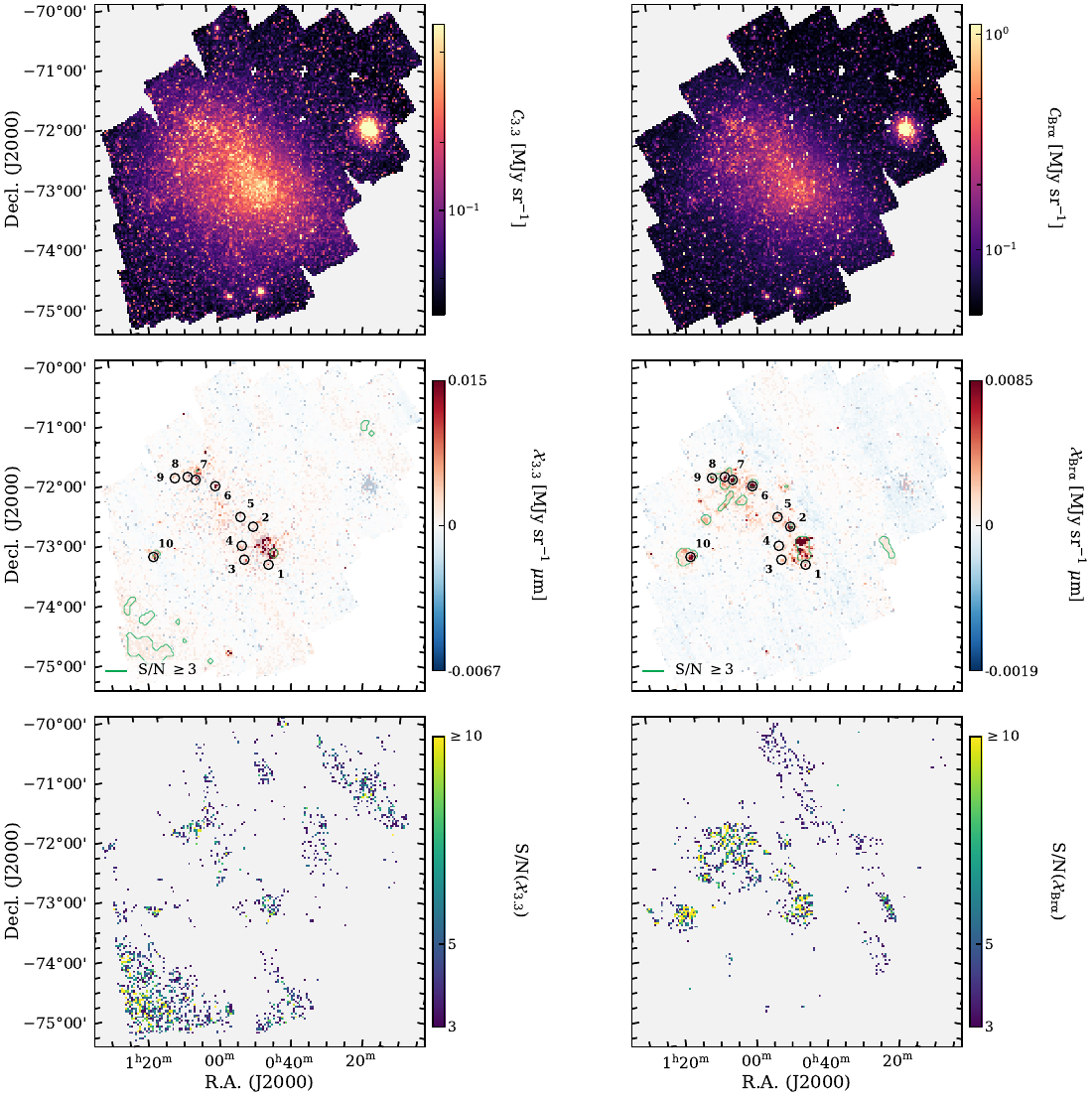}
\caption{Same as Figure~\ref{fig:lmc_full_spherex}, but for the SMC, where coherent \Xthree\ and \Xbralpha\ detections are more localized along the bar and active complexes.  Open circles with cloud-specific numeric IDs identify the SMC H{\sc ii} regions listed in Table~\ref{tab:hii_regions_individual}.}
\label{fig:smc_full_spherex}
\end{figure*}

\section{Methods}
\label{sec:methods}

\subsection{Matched Cells}

The fiducial cell grid uses 2 arcmin cells; a 4 arcmin version in the Appendix tests resolution dependence.  This scale reduces sensitivity to native-pixel noise while retaining environmental statistics in both Clouds.

\subsection{Fiducial Quality-Controlled Sample}

For quantitative fits, we require finite positive \Xthree, finite matched $\sigmad$ and TIR/$\sigmad$, full far-infrared coverage, and matched HERITAGE finite fractions of at least 0.75.  We do not impose a hard \Xthree\ S/N cut in the fiducial model because it would select on the dependent variable; S/N cuts are used only for display and robustness tests.

The fiducial 2 arcmin sample contains 12,882 LMC cells and 1,600 SMC cells, or 82\% and 71\% of the finite physical samples.  Both Clouds use the same cell definitions, quality criteria, and residual models, but sample different regimes: the LMC supplies most high-\Xbralpha\ active-region dynamic range, while the SMC contributes a lower-metallicity, lower-surface-brightness regime.

\subsection{Environment Definitions}

Cells are assigned to diffuse candidate, ionized non-H{\sc ii}, H{\sc ii} shell/near, or H{\sc ii} interior classes using H{\sc ii}-region geometry and H$\alpha$ context \citep{Gaustad2001,Lopez2014}.  With $d_{\rm HII}$ the distance to the nearest cataloged region and $R_{\rm HII}$ its radius, diffuse candidates have $d_{\rm HII}/R_{\rm HII}>5$ and no SHASSA H$\alpha$ excess above 300 dR; ionized non-H{\sc ii} cells have $d_{\rm HII}/R_{\rm HII}>5$ and H$\alpha\geq300$ dR; H{\sc ii} shell/near cells have $1<d_{\rm HII}/R_{\rm HII}\leq5$; and H{\sc ii} interiors have $d_{\rm HII}/R_{\rm HII}\leq1$.  We interpret these as quiescent diffuse cells, transition/ionized cells, and H{\sc ii} interiors.  The labels organize residuals by environment, where aromatic emission and PAH mass fraction are known to vary \citep{Gordon2008,Sandstrom2010,Oey2017,Chastenet2019,Chastenet2023PAHFraction,Egorov2023,Murgia2026}; they do not imply that a single physical mechanism operates in every cell.

\subsection{Column, Heating, and Residual Model}

For each Cloud we fit the quiescent diffuse baseline
\begin{equation}
\log \Xthree = a + b \log \sigmad + c \log\left(\frac{\tir}{\sigmad}\right),
\label{eq:baseline}
\end{equation}
where \Xthree\ is the 3.3 \um\ feature excess, $\sigmad$ is the dust surface density, and $\tir/\sigmad$ is the heating proxy.  This removes the first-order dust-column and radiation-field terms central to infrared dust-emission modeling \citep{DraineLi2007,Draine2007}.  We then define
\begin{equation}
\Deltathree = \log \Xthree - \log \Xthree(\sigmad,\tir/\sigmad).
\label{eq:residual}
\end{equation}
The coefficients are estimated by unweighted ordinary least squares in log space on the quiescent diffuse cells, separately for the LMC and SMC.  With these Cloud-specific predictions, \Deltathree\ measures whether cells are brighter or fainter than expected for their own Cloud's dust column and heating.  Appendix~\ref{app:product_validation} tests related heating proxies.

Residual-contrast uncertainties use a spatial block bootstrap: neighboring cells are grouped into 10--12 arcmin blocks, the blocks are resampled, and the diffuse baseline and regime medians are recomputed.  Quoted intervals therefore include spatial correlation and baseline-fit uncertainty.  When we summarize correlations between residuals and \Xbralpha, adjacent cells reduce the effective number of independent data points.  We therefore treat the Spearman rank coefficients, $r_{\rm S}$, as descriptive measures and quote bracketed 16--84 percentile ranges from the same spatial block resampling.

\subsection{Matched 8 Micron Aromatic-Band Control}

To test whether the 3.3 \um\ residual is selective or shared by aromatic-band-dominated emission, we repeat the diffuse-baseline decomposition for \Ieight.  For each Cloud we fit
\begin{equation}
\log \Ieight = a_8 + b_8 \log \sigmad + c_8 \log\left(\frac{\tir}{\sigmad}\right),
\label{eq:i8baseline}
\end{equation}
on quiescent diffuse cells with finite positive \Ieight, then define
\begin{equation}
\Deltaeight = \log \Ieight - \log \Ieight(\sigmad,\tir/\sigmad).
\label{eq:i8residual}
\end{equation}
The differential residual $\Deltathree-\Deltaeight$ measures whether the 3.3 \um\ feature is enhanced or deficient relative to 8 \um\ after the same dust-column and heating terms are removed.  The \Ieight\ decomposition uses the same cell grid and environment classes as \Xthree; requiring finite positive \Ieight\ removes no additional fiducial 2 arcmin \Xthree\ cells.

Because \Xthree\ is an integrated excess whereas \Ieight\ is a broadband intensity, Appendix~\ref{app:product_validation} also tests a SPHEREx 3.3 \um\ feature-window intensity,
\begin{equation}
\Ithreewin = \Cthree+\frac{\Xthree}{\Delta\lambda_{3.3}},
\label{eq:i33win}
\end{equation}
where $\Delta\lambda_{3.3}=0.22~\um$ is the width of the adopted 3.18--3.40 \um\ feature window.  \Ithreewin\ keeps the local continuum and is therefore closer in form to a broadband surface brightness, although much narrower than IRAC 8 \um.

\subsection{Local-Background H{\sc ii} Region Test}

To test cloud-scale gradients, we also compare H{\sc ii} interiors with clean 3--5 $R_{\rm HII}$ annuli.  This checks whether negative residuals persist relative to each region's surroundings, as in resolved aromatic/PDR studies \citep{Helou2004,Gordon2008,Murgia2026}.

\section{Results}
\label{sec:results}

\subsection{Full-Cloud Maps}

Figures~\ref{fig:lmc_full_spherex} and \ref{fig:smc_full_spherex} show the SPHEREx products used here.  \Xthree\ is widespread across both Clouds, whereas coherent \Xbralpha\ detections are more compact and localized.  The open circles and cloud-specific numeric IDs in the \Xthree\ and \Xbralpha\ panels identify the fiducial resolved H{\sc ii} regions used for the local-background test; Table~\ref{tab:hii_regions_individual} lists the corresponding catalog names, coordinates, radii, and local-background contrasts.  When we refer to active regions below, the classification comes from recombination-line emission and H{\sc ii}-catalog context, not from \Xthree\ brightness alone.  The aromatic-feature--recombination-line separation echoes resolved Spitzer and SPHEREx results in which aromatic emission is extended relative to hot dust or ionized gas \citep{Helou2004,Bendo2008,Gordon2008,Murgia2026}.  This motivates separating total feature brightness from residuals after dust column and heating are removed.

\begin{deluxetable*}{lrllcrrrrrr}
\tablecaption{Individual H{\sc ii}-region local-background residuals\label{tab:hii_regions_individual}}
\tablewidth{0pt}
\tablehead{
\colhead{Cloud} & \colhead{ID} & \colhead{Region} & \colhead{R.A.} & \colhead{Decl.} & \colhead{$R_{\rm HII}$} & \colhead{$N_{\rm int}$} & \colhead{$N_{\rm bg}$} & \colhead{$\Delta_{\rm loc}\log\mathcal{X}_{3.3}$} & \colhead{$\Delta_{\rm loc}\log\mathcal{X}_{\rm Br\alpha}$} & \colhead{$\Delta_{\rm loc}\Delta_{3.3}$} \\
\colhead{} & \colhead{} & \colhead{} & \colhead{(J2000)} & \colhead{(J2000)} & \colhead{(arcmin)} & \colhead{} & \colhead{} & \colhead{(dex)} & \colhead{(dex)} & \colhead{(dex)}
}
\startdata
LMC & 1 & N79 & 04:52:04 & -69:22:34 & 4.4 & 13 & 169 & +0.33 & +0.91 & -0.34 \\
LMC & 2 & N11 & 04:56:41 & -66:27:19 & 10.0 & 63 & 314 & +1.21 & +1.48 & -0.22 \\
LMC & 3 & N105 & 05:09:56 & -68:54:03 & 2.9 & 5 & 72 & +0.77 & +1.86 & -0.25 \\
LMC & 4 & N30 & 05:13:51 & -67:27:22 & 3.1 & 5 & 78 & +0.66 & +1.24 & -0.14 \\
LMC & 5 & N119 & 05:18:45 & -69:14:03 & 5.9 & 19 & 327 & +0.40 & +1.07 & -0.09 \\
LMC & 6 & N44 & 05:22:16 & -67:57:09 & 7.1 & 31 & 469 & +0.99 & +1.22 & -0.14 \\
LMC & 7 & N48 & 05:25:50 & -66:15:03 & 5.2 & 13 & 67 & +1.25 & +1.19 & +0.05 \\
LMC & 8 & N144 & 05:26:38 & -68:49:55 & 4.9 & 17 & 233 & +0.31 & +0.98 & -0.16 \\
LMC & 9 & N206 & 05:30:38 & -71:03:53 & 7.7 & 40 & 448 & +0.70 & +1.49 & -0.19 \\
LMC & 10 & N55 & 05:32:33 & -66:27:20 & 3.6 & 8 & 21 & +1.05 & +1.03 & -0.53 \\
LMC & 11 & N59 & 05:35:24 & -67:33:22 & 3.9 & 11 & 140 & +0.92 & +1.40 & -0.15 \\
LMC & 12 & N157 & 05:38:36 & -69:05:33 & 6.8 & 26 & 439 & +0.95 & +1.78 & -0.52 \\
LMC & 13 & N160 & 05:40:22 & -69:37:35 & 5.0 & 12 & 235 & +0.55 & +0.87 & -0.09 \\
LMC & 14 & N180 & 05:48:52 & -70:03:51 & 2.7 & 6 & 62 & +0.73 & +2.12 & -0.09 \\
SMC & 1 & N17 & 00:46:41 & -73:31:38 & 1.5 & 2 & 24 & +0.47 & +0.69 & +0.18 \\
SMC & 2 & N36 & 00:50:26 & -72:52:59 & 2.5 & 4 & 63 & +0.13 & +1.06 & -0.32 \\
SMC & 3 & N51 & 00:52:40 & -73:26:29 & 1.9 & 2 & 37 & +0.10 & +0.28 & -0.25 \\
SMC & 4 & DEM S74 & 00:53:14 & -73:12:18 & 2.7 & 6 & 70 & +0.44 & +0.24 & +0.04 \\
SMC & 5 & N50 & 00:53:26 & -72:42:56 & 4.3 & 10 & 159 & +0.01 & +0.65 & -0.20 \\
SMC & 6 & N66 & 00:59:06 & -72:10:44 & 3.6 & 2 & 122 & +0.06 & +0.94 & -0.28 \\
SMC & 7 & N76 & 01:03:32 & -72:03:16 & 3.1 & 8 & 91 & +0.40 & +0.77 & -0.01 \\
SMC & 8 & N78 & 01:05:18 & -71:59:53 & 2.6 & 5 & 67 & +0.83 & +1.13 & +0.01 \\
SMC & 9 & N80 & 01:08:13 & -72:00:06 & 2.2 & 4 & 35 & +0.70 & +0.94 & +0.21 \\
SMC & 10 & N84 & 01:14:56 & -73:17:51 & 5.7 & 18 & 64 & +1.08 & +1.27 & -0.02 \\
\enddata
\tablecomments{ID gives the numeric label used within each Cloud in the \Xthree\ and \Xbralpha\ panels of Figures~\ref{fig:lmc_full_spherex} and \ref{fig:smc_full_spherex}, and in the \Deltathree\ residual map in Figure~\ref{fig:delta33_residual_sky}.  Only fiducial resolved regions are listed: the clean 3--5 $R_{\rm HII}$ local-background annulus must contain at least five cells and the H{\sc ii}-region interior must contain at least two 2 arcmin cells.  $N_{\rm int}$ and $N_{\rm bg}$ are the number of finite cells entering the $\Delta_{3.3}$ local-background contrast.  The first two contrast columns show the absolute 3.3 $\mu$m feature and Br$\alpha$ enhancements relative to the same local annulus, while the last column gives the local contrast in the column+heating residual from Equation~(\ref{eq:residual}).}
\end{deluxetable*}

\subsection{Dust Column, Heating, and Environmental Residuals}

\begin{figure*}[!t]
\centering
\includegraphics[width=\textwidth]{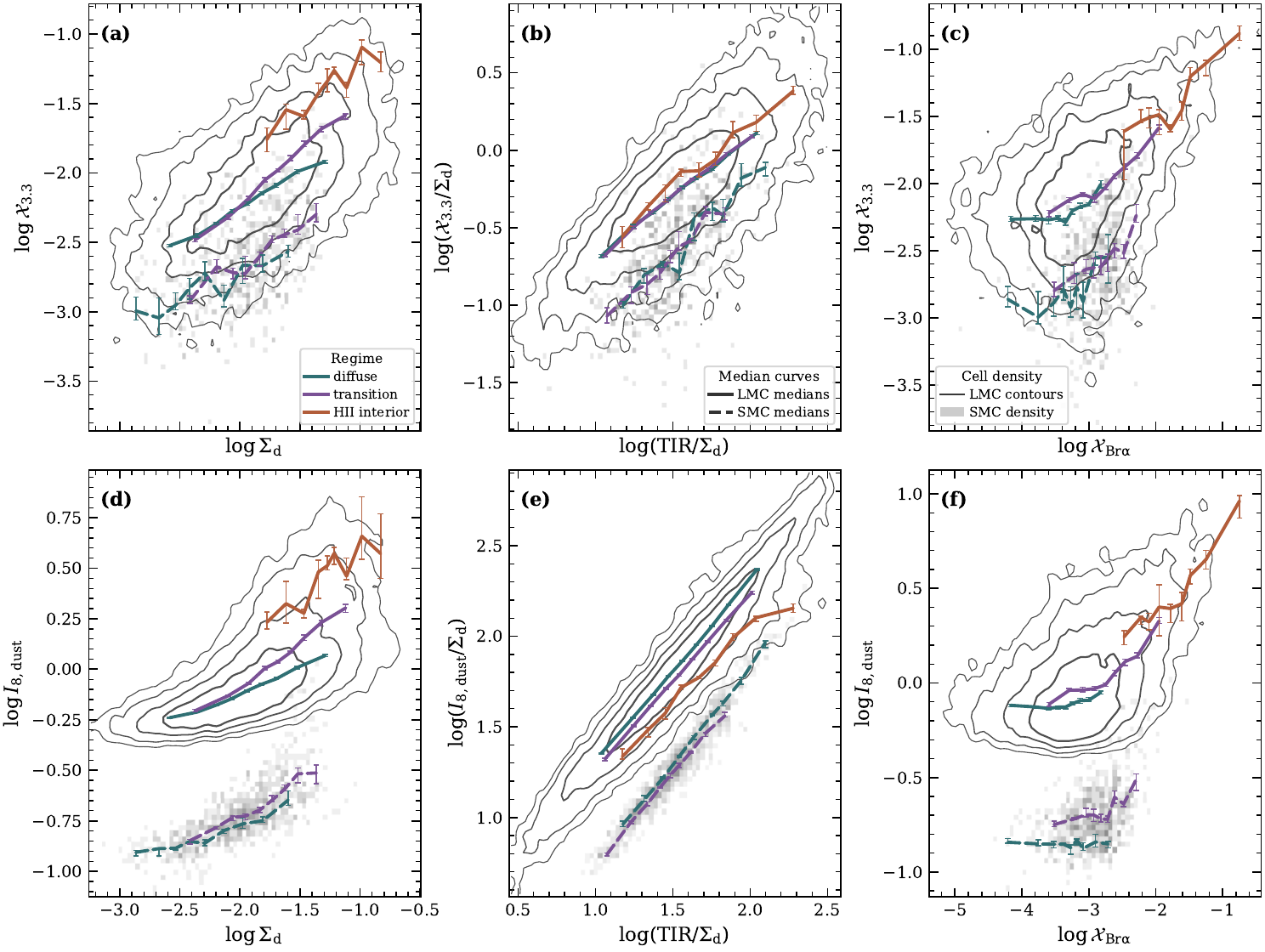}
\caption{Quality-controlled 2 arcmin relations for \Xthree\ (top) and \Ieight\ (bottom).  Columns compare intensity with $\sigmad$, intensity per dust mass with $\tir/\sigmad$, and intensity with \Xbralpha.  Gray-scale density shows SMC cells, contours show LMC cells, and colored curves show regime medians; solid and dashed curves denote the LMC and SMC.  Vertical bars give 16--84 percentile bootstrap uncertainties on the medians.}
\label{fig:physical_driver_scatter}
\end{figure*}

Figure~\ref{fig:physical_driver_scatter} compares \Xthree\ and \Ieight\ with dust column, heating, and Br$\alpha$ on the same cell grid and environment classes.  Both quantities rise with $\sigmad$, and both intensity-per-dust ratios rise with TIR/$\sigmad$, showing that dust column and radiative heating provide the first-order description.  Direct correlations with \Xbralpha\ are also positive, but they mix feature excitation, common star-forming structure, and residual environmental change.  The LMC reaches higher \Xthree, \Ieight, and \Xbralpha, while the SMC samples a fainter, lower-\Xbralpha\ regime.  The 8 \um\ relations are tighter, whereas \Xthree\ shows larger environmental separation, motivating diffuse-baseline fits for both observables.

\begin{figure*}[!t]
\centering
\includegraphics[width=\textwidth]{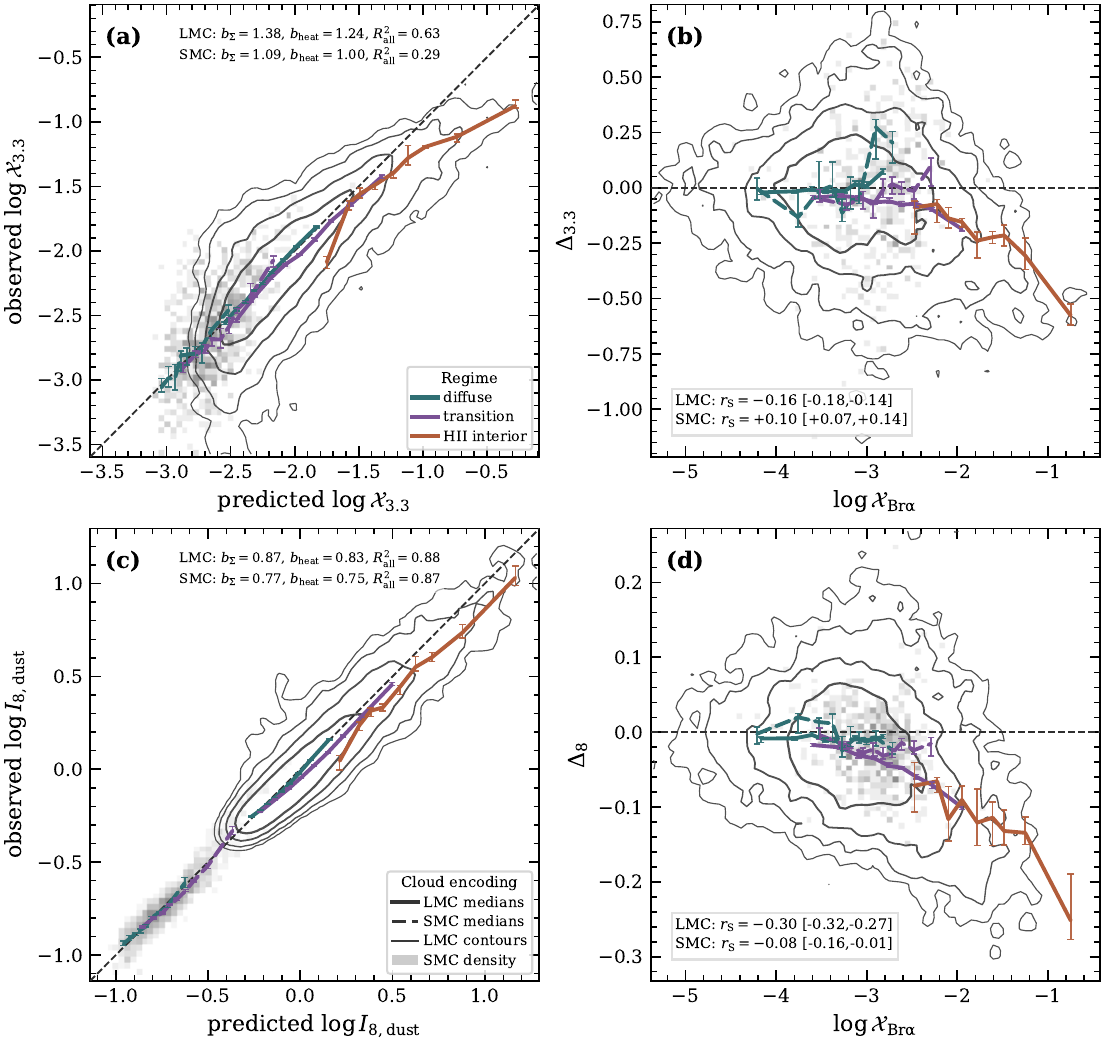}
\caption{Diffuse-baseline decompositions for \Xthree\ (top) and \Ieight\ (bottom).  Panels (a) and (c) compare observed intensities with quiescent diffuse predictions based on $\sigmad$ and $\tir/\sigmad$; panels (b) and (d) show residuals versus \Xbralpha.  Text annotations in panels (b) and (d) give descriptive Spearman rank coefficients, $r_{\rm S}$, with 16--84 percentile spatial block-bootstrap ranges.  Plotting conventions follow Figure~\ref{fig:physical_driver_scatter}.  Dashed black lines mark equality or zero residual.}
\label{fig:baseline_residual_comparison}
\end{figure*}

Figure~\ref{fig:baseline_residual_comparison} applies the diffuse-baseline decomposition in parallel.  Dust column and heating predict \Ieight\ more tightly than \Xthree: all-cell $R^2$ values are 0.881 and 0.866 for \Ieight\ in the LMC and SMC, while the corresponding \Xthree\ values are lower (Table~\ref{tab:baseline_fit_coefficients}).  High-\Xbralpha\ LMC H{\sc ii}-interior cells lie below zero for both residuals, but more strongly for \Deltathree\ than for \Deltaeight.  SMC residual sequences lie closer to zero over the lower-\Xbralpha\ range sampled there.  Thus Br$\alpha$ is not the primary predictor of total \Xthree, but marks where residual departures from the diffuse column+heating expectation occur.

For the \Xthree\ diffuse baseline, Table~\ref{tab:baseline_fit_coefficients} lists the fiducial 2 arcmin coefficients, fit statistics, and 16--84 percentile spatial block-bootstrap intervals.  The LMC fit uses 7026 diffuse cells from 12882 coverage-safe cells and 974 spatial blocks, giving $R^2_{\rm diff}=0.547$ and $R^2_{\rm all}=0.628$.  The SMC fit uses 492 diffuse cells from 1600 coverage-safe cells and 141 blocks, giving $R^2_{\rm diff}=0.137$ and $R^2_{\rm all}=0.288$.  Because $R^2_{\rm all}$ is evaluated on the broader coverage-safe sample, which spans a larger range in dust column and heating than the diffuse subset, it can exceed $R^2_{\rm diff}$ even though the coefficients are fitted only to diffuse cells; it is a predictive summary, not a refit to all cells.  The lower SMC fit quality is not due to a different estimator; it reflects the smaller, lower-dynamic-range SMC sample and its larger diffuse-fit scatter, so the broader SMC baseline uncertainty is propagated into the residual bootstrap.

\begin{table*}[!t]
\caption{Diffuse-baseline coefficient uncertainties for \Xthree}
\label{tab:baseline_fit_coefficients}
\centering
\scriptsize
\begin{tabular}{@{}lcccccc@{}}
\hline\hline
Cloud & $a$ & $b_\Sigma$ & $c_{\rm heat}$ & $R^2_{\rm diff}$ & $R^2_{\rm all}$ & RMSE \\
 & & & & & & (dex) \\
\hline
LMC & $-1.468\,[-1.484,-1.452]$ & $1.377\,[1.353,1.399]$ & $1.242\,[1.210,1.271]$ & $0.547\,[0.527,0.567]$ & $0.628\,[0.615,0.642]$ & $0.241\,[0.234,0.247]$ \\
SMC & $-1.993\,[-2.104,-1.881]$ & $1.091\,[0.931,1.249]$ & $1.003\,[0.779,1.232]$ & $0.137\,[0.108,0.171]$ & $0.288\,[0.231,0.323]$ & $0.484\,[0.456,0.509]$ \\
\hline
\end{tabular}
\vspace{0.5ex}
\begin{minipage}{0.98\textwidth}
\scriptsize
\textit{Note--} Coefficients are for the fiducial 2 arcmin model $\log\Xthree=a+b_\Sigma\log\sigmad+c_{\rm heat}\log(\tir/\sigmad)$, fit by unweighted ordinary least squares on quiescent diffuse cells separately in each Cloud. The numbers of all cells, diffuse fit cells, and spatial blocks are given in the text. Bracketed ranges are 16--84 percentile intervals from 1000 spatial block-bootstrap draws using 10 arcmin blocks; the baseline is refit in every draw. $R^2_{\rm diff}$ is evaluated only on the diffuse cells used for fitting, while $R^2_{\rm all}$ evaluates the same diffuse baseline on all coverage-safe cells.
\end{minipage}
\end{table*}

Relative to the quiescent diffuse baseline, LMC transition/ionized cells are depressed by $-0.076^{+0.009}_{-0.009}$ dex and H{\sc ii} interiors by $-0.223^{+0.035}_{-0.040}$ dex.  The corresponding SMC medians are $-0.029^{+0.032}_{-0.041}$ dex and $-0.011^{+0.087}_{-0.126}$ dex, consistent with zero over the lower-\Xbralpha\ range sampled there.  Appendix Figure~\ref{fig:robustness_summary} shows that the LMC active-region residual remains negative at 4 arcmin and under coverage variants, while SMC intervals remain broader and centered near zero.

The environmental signal is not that H{\sc ii} regions are absolutely faint, but that they are under-luminous relative to the \Xthree\ expected from their dust column and heating.  Because the free-free curvature residual discussed in Section~\ref{sec:data} has positive sign, correcting for it would make the active-region 3.3 \um\ deficits slightly stronger.

\begin{figure*}[t]
\centering
\includegraphics[width=0.98\textwidth]{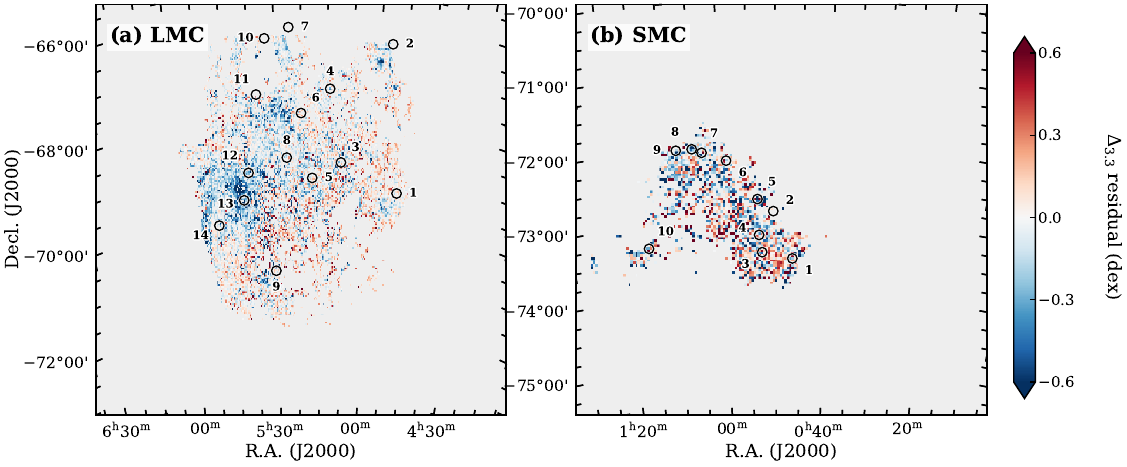}
\caption{Sky distribution of the diffuse-baseline residual \Deltathree\ for coverage-safe 2 arcmin cells in the LMC and SMC.  Blue cells are under-luminous and red cells are over-luminous relative to the quiescent diffuse expectation at fixed dust column and TIR/$\sigmad$; the color scale is clipped at $|\Deltathree|=0.6$ dex for display.  Open circles and numeric IDs mark the fiducial resolved H{\sc ii} regions listed in Table~\ref{tab:hii_regions_individual}, using the same marker convention as Figures~\ref{fig:lmc_full_spherex} and \ref{fig:smc_full_spherex}.}
\label{fig:delta33_residual_sky}
\end{figure*}

Figure~\ref{fig:delta33_residual_sky} maps these \Deltathree\ residuals for the same coverage-safe 2 arcmin cells.  In the LMC, negative residuals form coherent structures around several active complexes rather than being confined to one extreme object; this map-space pattern is consistent with the individual-region result that 13 of 14 fiducial LMC H{\sc ii} regions have negative local \Deltathree\ contrasts.  The SMC residual map is more mixed over its lower-\Xbralpha\ dynamic range, with six of ten fiducial regions showing negative local contrasts and an aggregate contrast consistent with zero.

\subsection[Residual Comparison between 3.3 and 8 Micron Emission]{Residual Comparison between 3.3 and 8 Micron Emission}
\label{sec:selective_residual}

With the parallel decompositions established, \Ieight\ becomes a matched aromatic-band control: do active regions show a generic aromatic-band-dominated residual deficit, or is the deficit stronger in the isolated 3.3 \um\ feature?  Adding Br$\alpha$ after column and heating changes the \Ieight\ prediction only weakly, so \Ieight\ mainly tests the selectivity of the \Xthree\ residual.

\begin{figure}[t]
\centering
\includegraphics[width=0.45\textwidth]{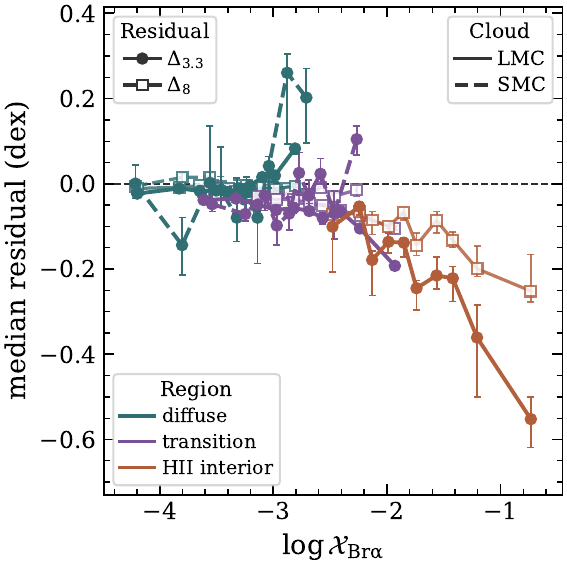}
\caption{Common Br$\alpha$ residual sequence for \Xthree\ and \Ieight, combining panels (b) and (d) of Figure~\ref{fig:baseline_residual_comparison}.  Colors identify region type; solid and dashed lines denote the LMC and SMC; filled circles and open squares show \Deltathree\ and \Deltaeight.  The SMC samples lower \Xbralpha, while LMC H{\sc ii} interiors reach the highest \Xbralpha\ and show a stronger negative \Deltathree\ trend than \Deltaeight.}
\label{fig:bralpha_residual_sequence}
\end{figure}

Figure~\ref{fig:bralpha_residual_sequence} overlays the median \Deltathree\ and \Deltaeight\ sequences.  SMC tracks populate the lower-\Xbralpha\ regime with residuals consistent with zero within broader intervals; LMC tracks extend to higher-\Xbralpha\ active regions where both residuals become negative.

The 8 \um\ residual sequence is milder.  LMC H{\sc ii} interiors have median $\Deltaeight=-0.102$ dex, compared with $\Deltathree=-0.223$ dex, giving $\Deltathree-\Deltaeight=-0.122$ dex.  Transition/ionized cells have the same sign but smaller amplitude, and SMC medians are consistent with zero at lower \Xbralpha.  The active-region deficit is therefore not a generic reduction of all aromatic-band-dominated infrared emission, but is strongest in the small-carrier-sensitive 3.3 \um\ feature.

Appendix~\ref{app:product_validation} shows that this high-\Xbralpha\ differential result is insensitive to plausible low-\Ieight\ cuts: the LMC H{\sc ii}-interior $\Deltathree-\Deltaeight$ remains negative, from $-0.122$ to $-0.053$ dex.  The same cuts leave SMC cells in the lower-\Xbralpha\ regime with weak differential residuals.

The result also survives the definition-control test: using \Ithreewin\ gives an LMC H{\sc ii}-interior contrast of $-0.296$ dex, compared with $-0.223$ dex for \Xthree\ and $-0.102$ dex for \Ieight.  A broadband-like 3.3 \um\ quantity therefore remains more under-luminous in active LMC regions than \Ieight.

\subsection[Local-Background HII Region Measurements]{Local-Background H{\sc ii} Region Measurements}

Table~\ref{tab:hii_regions_individual} gives the individual-region measurements, and Table~\ref{tab:hii_local} summarizes the aggregate local-background contrasts.  The LMC residual contrast is not driven by a single object: 13 of 14 fiducial LMC regions have negative local \Deltathree\ contrasts, while all are bright in absolute \Xthree\ and \Xbralpha\ relative to their clean 3--5 $R_{\rm HII}$ annuli.  In aggregate, LMC H{\sc ii} interiors are enhanced by $+0.749$ dex in \Xthree\ and $+1.226$ dex in \Xbralpha, yet their column+heating residual is $-0.153^{+0.014}_{-0.043}$ dex.  The 10-region SMC measurement also shows enhanced \Xthree\ and \Xbralpha; six of ten regions have negative local \Deltathree\ contrasts, while the aggregate residual contrast, $-0.019^{+0.033}_{-0.184}$ dex, remains consistent with zero within broad intervals.

\begin{table}[!htbp]
\caption{Fiducial local-background H{\sc ii}-interior contrasts}
\label{tab:hii_local}
\centering
\scriptsize
\setlength{\tabcolsep}{1.8pt}
\begin{tabular}{@{}llccc@{}}
\hline
Cloud & Quantity & $N$ & $\Delta_{\rm loc}$ & $f_{<0}$ \\
\hline
LMC & $\mathcal{X}_{3.3}$ & 14 & $+0.749^{+0.169}_{-0.069}$ & 0.00 \\
LMC & $\mathcal{X}_{\rm Br\alpha}$ & 14 & $+1.226^{+0.172}_{-0.083}$ & 0.00 \\
LMC & $\mathcal{X}_{3.3}/\Sigma_{\rm d}$ & 14 & $+0.204^{+0.041}_{-0.122}$ & 0.36 \\
LMC & \Deltathree & 14 & $-0.153^{+0.014}_{-0.043}$ & 0.93 \\
SMC & $\mathcal{X}_{3.3}$ & 10 & $+0.417^{+0.053}_{-0.166}$ & 0.00 \\
SMC & $\mathcal{X}_{\rm Br\alpha}$ & 10 & $+0.856^{+0.087}_{-0.145}$ & 0.00 \\
SMC & $\mathcal{X}_{3.3}/\Sigma_{\rm d}$ & 10 & $+0.098^{+0.049}_{-0.087}$ & 0.40 \\
SMC & \Deltathree & 10 & $-0.019^{+0.033}_{-0.184}$ & 0.60 \\
\hline
\end{tabular}
\tablecomments{$N$ is the number of H{\sc ii} regions.  $\Delta_{\rm loc}$ is the median dex contrast relative to each region's clean 3--5 $R_{\rm HII}$ local background, and $f_{<0}$ is the fraction of regions with negative contrast.  \Deltathree\ denotes the column+heating residual from Equation~(\ref{eq:residual}).  Uncertainties give the 16--84 percentile bootstrap interval over regions.}
\end{table}

Thus H{\sc ii} regions can be bright 3.3 \um\ emitters because excitation is high, while still being deficient after dust column and heating are accounted for.

\section{Discussion}
\label{sec:discussion}

\subsection{Carrier Interpretation of the 3.3 Micron Feature}

The 3.3 \um\ UIB is generally assigned to aromatic C--H stretching and is often modeled with PAH populations, representing an ensemble of PAH-like aromatic carbonaceous carriers rather than an identified molecular species \citep{LegerPuget1984,Allamandola1985,Allamandola1989,Tielens2008,Peeters2021}.  This broad interpretation has been present since the early observational literature: the feature was recognized as unidentified, likely carbonaceous, and difficult to assign to a single terrestrial or laboratory analogue \citep{RussellSoiferMerrill1977,Knacke1977,WillnerPuetterRussellSoifer1979,TokunagaSellgrenSmith1991}.

Several carrier classes remain relevant to the interpretation of a SPHEREx 3.3 \um\ excess map.  In addition to gas-phase PAH molecules and ions, the literature includes PAH clusters or evaporating very small grains \citep{Rapacioli2005,Pilleri2012}, hydrogenated amorphous carbon and related amorphous carbon solids \citep{DuleyWilliams1981,ScottDuleyJahani1997,DuleyWilliams2011}, quenched carbonaceous composite materials \citep{Sakata1990}, coal/kerogen-like organics \citep{Papoular1989,Papoular2001}, and mixed aromatic--aliphatic organic nanoparticles \citep{KwokZhang2011,KwokZhang2013,Kwok2022}.  The 3.3--3.4 \um\ subfeatures and recent JWST/AKARI measurements further show that aromatic, olefinic, aliphatic, sidegroup, hydrogenation, deuteration, and anharmonic effects can change the detailed spectrum \citep{vanDiedenhoven2004,Yang2013,Yang2016,Yang2017,Sadjadi2017,Peng2026Dor,Peng2026AKARI}.  The 3.3 \um\ window also contains H{\sc i} Pf$\delta$ at 3.297 \um.  This matters because the feature carrier remains debated \citep{TokunagaBernstein2021}, and because atomic-hydrogen \citep{Zagury2021,Zagury2023} and Rydberg-matter \citep{BadieiHolmlid2002,Holmlid2003} alternatives have been discussed for UIB-related emission.  Section~\ref{sec:hydrogen_origin_checks} therefore tests Br$\alpha$-scaled H{\sc i} contamination and raw D4 stacks directly.  The \Xthree\ residuals are best interpreted as feature power per dust-column/heating expectation, rather than as the abundance of one PAH species.

We therefore use \Xthree\ and \Deltathree\ as empirical measures of 3.3 \um\ feature power after dust column and heating are accounted for.  They do not by themselves distinguish carrier abundance, emissivity, bonding state, charge, hydrogenation, or geometry.  A positive total \Xthree\ in an H{\sc ii} region can still coexist with a negative residual relative to the diffuse baseline.

\subsection[H I and Rydberg-Related Origin Checks]{H{\sc i} and Rydberg-Related Origin Checks}
\label{sec:hydrogen_origin_checks}

We also test whether H{\sc i} Pf$\delta$ or related hydrogen/Rydberg-line emission could dominate the SPHEREx 3.3 \um\ feature measurement.  Using Case-B emissivities, Pf$\delta$/Br$\alpha$ = 0.0933 \citep{HummerStorey1987}.  The direct Pf$\delta$ fraction of \Xthree\ is small: median values are 0.004, 0.014, and 0.049 for LMC diffuse, transition, and H{\sc ii}-interior cells, and 0.014, 0.041, and 0.083 for the corresponding SMC classes.  Subtracting this Case-B component, refitting the diffuse baseline, and recomputing residuals makes the LMC H{\sc ii}-interior contrast more negative, from $-0.223$ to $-0.247$ dex; the SMC H{\sc ii}-interior contrast remains consistent with zero, changing from $-0.011$ to $-0.060$ dex with a 16--84 percentile range of $-0.149$ to $+0.009$ dex.  Thus ordinary Case-B Pf$\delta$ contamination cannot create the observed active-region 3.3 \um\ deficit; if present, it dilutes the deficit.

The continuum sidebands make the net H{\sc i} contribution smaller than direct Pf$\delta$ alone.  Including Pf$\gamma$ in the red continuum sideband gives median net H{\sc i} fractions of 0.024 and 0.040 in LMC and SMC H{\sc ii}-interior cells, respectively, and adding a plausible Pf$\epsilon$ contribution in the blue sideband drives the median net contribution close to zero.  Independent H{\sc i} line-window stacks give the same conclusion: the Pf$\delta$+3.3 \um\ window is much brighter than Case B, while the nearby Pf$\gamma$ window does not show a comparable excess.  A hydrogen-line interpretation strong enough to dominate \Xthree\ would therefore require line-ladder behavior that is not seen in the neighboring SPHEREx windows.

The raw D4 stacks provide a complementary profile check.  Using high-S/N cells, 0.025 \um\ bins, and the calibrated sample wavelengths, a broad 3.3 \um\ template is preferred over an unresolved Pf$\delta$-like template in five of six cloud/environment stacks, although the two templates remain highly correlated at SPHEREx resolution.  A three-component decomposition with fixed Case-B Pf$\delta$, a main 3.3 \um\ component, and a broad 3.4 \um\ satellite/plateau component gives H{\sc ii}-interior Pf$\delta$ fractions of only 0.071 in the LMC and 0.078 in the SMC, while the red 3.4/3.3 ratios are 0.46 and 0.22.  SPHEREx therefore cannot identify the microscopic carrier uniquely, but the measured feature does not behave like a Br$\alpha$-scaled H{\sc i} recombination component.

\subsection{Physical Interpretation and Relation to Previous Work}

Given this carrier scope and the H{\sc i} checks above, the environmental signal is interpreted as a differential residual rather than as a direct PAH-abundance measurement.  Active regions can be bright in total \Xthree\ because dust column and heating are high, yet under-luminous relative to the diffuse baseline.  This is plausible because aromatic-band emission depends on grain size, charge, hydrogenation, excitation, and survival in UV fields, shocks, and hot gas \citep{Allain1996a,Allain1996b,Tielens2008,Micelotta2010a,Micelotta2010b,Rigopoulou2021}.  The residual traces feature power relative to the diffuse dust-column/heating expectation and is distinct from the dust-model PAH mass fraction, $\qpah$; the 3.3 \um\ band is weighted toward small aromatic grains \citep{DraineLi2007,Draine2021,Lai2020,Rigopoulou2021,Whitcomb2024,Whitcomb2026}.

Previous work has shown that aromatic emission, dust-model PAH fractions, and aromatic/UIB band ratios vary with metallicity and active-region environment \citep{Gordon2008,Sandstrom2010,Sandstrom2012,Oey2017,Chastenet2019,Vermeij2002,Mori2012,Whitcomb2024,Whitcomb2026,Koziol2026}, and JWST/SPHEREx now resolve aromatic-feature/recombination-line structure around ionized gas \citep{Sandstrom2023PAH3,Chastenet2023PAHFraction,Egorov2023,Pedrini2024,Gregg2024,Murgia2026}.  In early SPHEREx maps of Cygnus X and the North American Nebula, \citet{Hora2026} found that 3.28 \um\ emission broadly follows 7.7 and 11.2 \um\ aromatic-band emission while retaining smaller differences that may trace grain-size and UV-field variations.  \citet{Boersma2026} combined SPHEREx and Spitzer maps of the NGC~7023 PDR to separate the 3.3 and 3.4 \um\ subfeatures and found that PAH band ratios and PAHdb-derived size measures change between dense and diffuse structures.  The contribution here is the two-Cloud matched-cell decomposition of \Xthree\ against dust column, heating, Br$\alpha$, and \Ieight.  The SMC constrains the lower-\Xbralpha, lower-surface-brightness regime, while the LMC provides high-\Xbralpha\ leverage where the selective 3.3 \um\ residual deficit is clearest.  The stronger LMC deficit at 3.3 \um\ than at 8 \um\ points to selective changes in the 3.3 \um-emitting carrier population, plausibly through preferential destruction or transformation of the smallest aromatic/aliphatic carbonaceous carriers, altered charge or hydrogenation, or geometric dilution inside ionized regions.

\subsection{Connection to the 2175 \texorpdfstring{\AA}{Angstrom} Bump}

A similar environmental theme appears in studies of the 2175~\AA\ ultraviolet bump, another diagnostic associated with small carbonaceous or aromatic material.  Resolved attenuation-curve work finds that the UV bump weakens with increasing H$\alpha$ surface brightness in both star-forming and non-star-forming regions \citep{Zhou2023,Guo2025,Guo2026Bump}.  Dust-model interpretations point to changes in the small carbonaceous grain population \citep{Guo2026DustModel}.  The connection is phenomenological, but it reinforces a common picture in which active environments can process small carbonaceous dust in ways hidden by total infrared brightness.

\subsection{Implications for Integrated and High-Redshift Measurements}

The 3.3 \um\ feature is increasingly used as a star-formation or ISM diagnostic in nearby and high-redshift galaxies \citep[e.g.,][]{Kim2012,Lai2020,Rigopoulou2021,Spilker2023,Zhang2025,Cheng2025,Lee2025,Whitcomb2024,Whitcomb2026}.  The Magellanic Clouds show why it should be interpreted with care: high integrated emission can reflect high dust column and excitation, while low emission per unit dust or TIR can indicate aromatic-carrier processing, dilution, or a changed small-grain population.

\subsection{Limitations}

This work depends on matched-resolution products, dust-model systematics, the TIR/$\sigmad$ heating proxy, and SPHEREx continuum windows.  The diffuse baseline is empirical rather than a full dust-emission model; unmodeled radiation-field hardness, geometry, dust-temperature structure, or dust-model biases can contribute to residuals.  Bootstrap intervals quantify spatially correlated median contrasts under the adopted model, not a full hierarchical error budget.  The \Ieight\ comparison is a relative aromatic-band control, and the residual remains phenomenological: it can reflect carrier destruction or transformation, altered size or charge, geometry, or heating-proxy errors \citep{DraineLi2007,Allain1996a,Allain1996b,Micelotta2010a,Micelotta2010b,Whitcomb2024}.  The SMC is essential to the two-Cloud comparison, but fewer retained cells, lower surface brightness, and smaller high-\Xbralpha\ dynamic range mean that its non-detection of strong residual deficits constrains the lower-activity regime rather than identical environments.

\section{Summary}
\label{sec:summary}

We combined SPHEREx 3.3 \um\ aromatic-feature and Br$\alpha$ maps with SAGE 8 \um\ emission, HERITAGE dust columns, a far-infrared heating proxy, and ionized-gas environments.  The main conclusions are:

\begin{enumerate}
\item \Xthree\ is widespread across both Clouds and is not simply an ionized-gas map.  Case-B Pf$\delta$ subtraction, sideband-aware H{\sc i} estimates, and raw D4 stacks show that ordinary hydrogen recombination cannot dominate the measured feature.  A quiescent diffuse column+heating baseline predicts the 2 arcmin sample with all-cell $R^2=0.628$ in the LMC and 0.288 in the SMC.
\item \Xbralpha\ is not the dominant predictor of total \Xthree\ after column and heating are included, but it identifies where negative residuals occur.  LMC H{\sc ii} interiors are under-luminous by $-0.223$ dex at 2 arcmin and remain negative at 4 arcmin; SMC H{\sc ii}/transition residuals are consistent with zero over the lower-\Xbralpha, lower-surface-brightness regime sampled there.
\item Compared with \Ieight, \Xthree\ is less completely predicted by dust column and heating and shows the stronger LMC H{\sc ii}-interior deficit, implying selective processing of the 3.3 \um-emitting small aromatic-carrier population rather than a uniform deficit in all aromatic-band-dominated infrared emission.
\item H{\sc ii} regions are bright in absolute \Xthree\ but deficient relative to their dust-column and heating expectation, separating feature excitation from environmental processing, dilution, or small-grain-population changes in resolved low-metallicity environments.
\end{enumerate}

\section*{Data and Software Availability}

The SPIRAL pipeline is publicly available.  The versioned SPIRAL v0.1.0 software archive used for this paper is available from Zenodo: \doi{10.5281/zenodo.22103715}.  The repository is maintained at \url{https://github.com/cli-tsinghua/SPIRAL}, which provides the code and documentation needed to reproduce the LMC and SMC maps from public SPHEREx L2 products.  Users of SPIRAL should cite the versioned software archive and this paper.  The public input products are available from their source archives: \dataset[SPHEREx QR2]{\doi{10.26131/IRSA652}}, Spitzer \dataset[SAGE]{\doi{10.26131/IRSA404}} and \dataset[SAGE-SMC]{\doi{10.26131/IRSA431}}, Herschel \dataset[HERITAGE]{\doi{10.26131/IRSA76}}, \dataset[SHASSA]{https://amundsen.swarthmore.edu/SHASSA/} H$\alpha$ maps, the \dataset[AllWISE Source Catalog]{\doi{10.26131/IRSA1}}, and the \dataset[2MASS Point Source Catalog]{\doi{10.26131/IRSA2}}.  The derived SPHEREx maps and manuscript-level measurement tables are available from Zenodo: \dataset[Derived SPHEREx LMC/SMC maps and measurements]{\doi{10.5281/zenodo.22103090}}.

\begin{acknowledgments}
The author thanks the anonymous referee for constructive comments and careful review.
This work is supported by the National Key R\&D Program of China (grant No. 2018YFA0404502), and the National Science Foundation of China (grant No. 12433003). This publication makes use of data products from the Spectro-Photometer for the History of the Universe, Epoch of Reionization and Ices Explorer (SPHEREx), which is a joint project of the Jet Propulsion Laboratory and the California Institute of Technology, and is funded by the National Aeronautics and Space Administration.  This work also uses archival products from the Spitzer SAGE and SAGE-SMC surveys and the Herschel HERITAGE survey, and from SHASSA, WISE, and 2MASS.
\end{acknowledgments}

\facilities{SPHEREx, Spitzer, Herschel, SHASSA, WISE, 2MASS}

\software{Astropy \citep{Astropy2013,Astropy2018,Astropy2022}, Matplotlib \citep{Hunter2007}, NumPy \citep{Harris2020}, pandas \citep{pandas2020,McKinney2010}, SciPy \citep{Virtanen2020}, SPIRAL \citep{SPIRALZenodo2026}}

\appendix

\section{SPHEREx Product and Robustness Checks}
\label{app:product_validation}

All figures and tables use quality-masked, wavelength-registered SPHEREx products from the pipeline described in Section~\ref{sec:data}.  Table~\ref{tab:validation_summary} summarizes the validation ladder: algorithmic checks, aperture closure, injection/recovery, negative-control windows, split-sample repeatability, pipeline variants, and external photometric comparisons.  These tests verify wavelength-window selection, zodiacal subtraction, masking, map-making, continuum fitting, and feature-flux preservation.  WISE and 2MASS comparisons constrain astrometry, morphology, and continuum scale \citep{Skrutskie2006,Wright2010}.

\begin{table}
\caption{Quantitative validation summary for the SPHEREx products}
\label{tab:validation_summary}
\centering
\scriptsize
\begin{tabular}{p{0.17\textwidth}p{0.24\textwidth}p{0.31\textwidth}p{0.20\textwidth}}
\hline
Validation component & Sample & Quantitative result & Contribution to uncertainty assessment \\
\hline
IRSA mosaic comparison & Five LMC/SMC 3.3 $\mu$m and Br$\alpha$ fields & $r=0.995\mbox{--}1.000$; median residual $-0.23\mbox{--}0.20\%$; NMAD $0.35\mbox{--}3.97\%$; best shift $(0,0)$ & External morphology, registration, and mean-scale reference \\
WISE/2MASS point sources & Clean catalog stars in representative LMC tiles & W1: $N=1328$, $r=0.93$, ratio=0.98; H: $N=85$, $r=0.96$, ratio=0.96; $K_s$: $N=139$, $r=0.98$, ratio=0.94 & Independent astrometric and continuum flux-scale constraint \\
Aperture closure & Direct raw-cutout spectra versus final map cells & $|\tilde\Delta \mathcal{X}|=0.0015$ MJy sr$^{-1}$ $\mu$m; NMAD $\leq 0.0030$ & Map-making preservation of aperture-level feature flux \\
Injection/recovery & Artificial continua and line/PAH excesses in raw cutouts & line/excess ratio $0.985\mbox{--}0.991$; continuum ratio $0.985\mbox{--}0.988$ & Feature-window measurement bias \\
Negative-control windows & Nearby line-free windows & local-window $|\tilde{\mathcal{X}}|=(3.2\mbox{--}4.0)\times10^{-5}$; $p_{84}\leq 9.3\times10^{-5}$ & Continuum-fit false-excess level \\
Split-null repeatability & Independent observation/time/version subsets & full-depth NMAD $p_{84}$ $3.6\times10^{-4}\mbox{--}1.7\times10^{-3}$ & Repeatability and empirical random floor \\
Pipeline variants & Coaddition weights and feature/continuum windows & coadd $p_{84}$ fractional shift $\leq 0.24$; window $p_{84}$ fractional change $\leq 0.27$ & Sets systematic floor for amplitude-sensitive conclusions \\
\hline
\end{tabular}
\end{table}

\begin{figure*}[!t]
\centering
\includegraphics[width=0.98\textwidth]{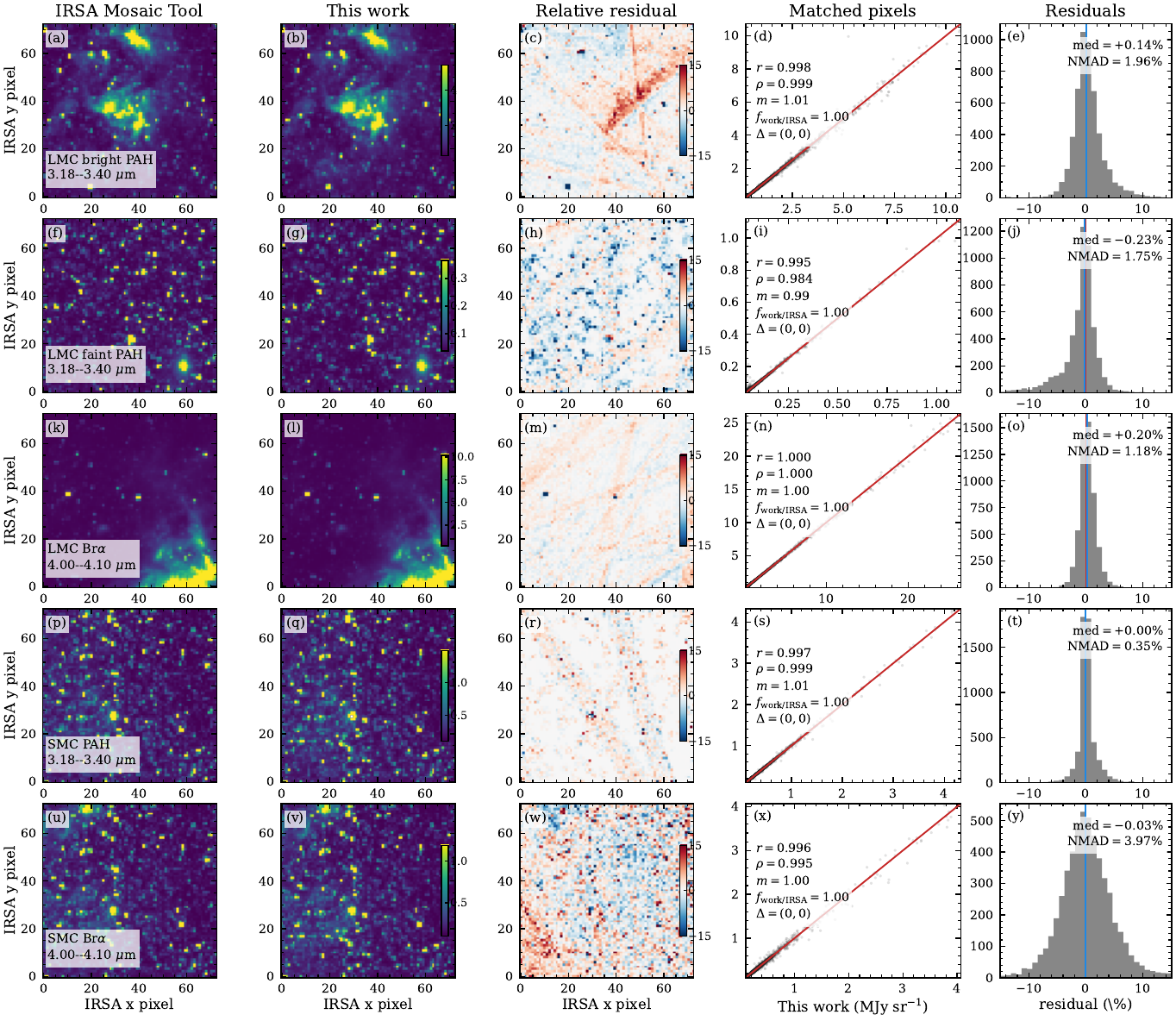}
\caption{External comparison with IRSA SPHEREx Mosaic Tool products.  Rows show five 3.3 \um\ or Br$\alpha$ validation fields; columns show the IRSA mosaic, our product on the IRSA grid, relative residual $100({\rm product}/{\rm IRSA}-1)$, matched-pixel comparison, and residual histogram.  Row maps use the same linear intensity stretch.  Annotations give log-space correlations, fitted slope, median product/IRSA ratio, and best integer-pixel shift.}
\label{fig:irsa_benchmark}
\end{figure*}

Figure~\ref{fig:irsa_benchmark} compares our products with IRSA SPHEREx Mosaic Tool products in five fields.  The IRSA products provide an external reference for morphology, registration, and mean flux scale, but not an absolute calibration standard.  With identical tabulated channel windows, the products show consistent morphology, median flux scale, and no preferred integer-pixel astrometric shift.  The IRSA residual NMAD range of $0.35\mbox{--}3.97\%$ is much smaller than the primary LMC H{\sc ii}-interior residual contrast of $-0.223$ dex, corresponding to a $\simeq40\%$ under-luminosity.  Weak residual texture follows coverage geometry and structured-emission gradients, as expected for early SPHEREx diffuse-emission mosaics \citep{Bock2026,Hui2026,Cukierman2026,Murgia2026}, and is treated as a systematic floor.

\begin{figure*}[!t]
\centering
\includegraphics[width=0.98\textwidth]{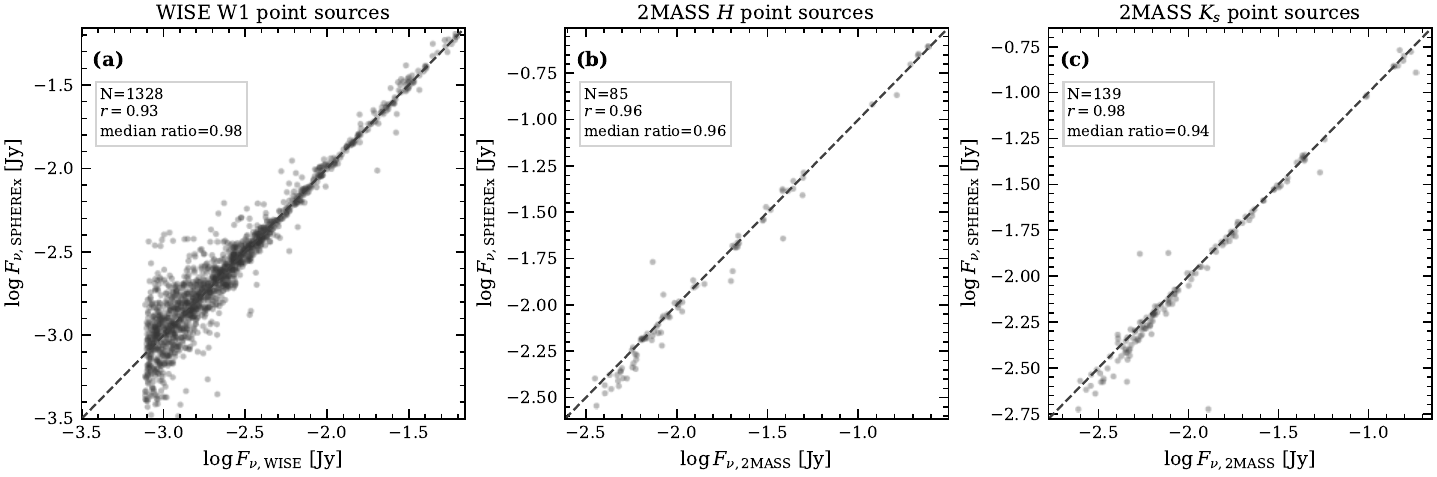}
\caption{Independent point-source validation using WISE and 2MASS data.  Panels compare SPHEREx synthetic aperture photometry of clean point sources with AllWISE W1, 2MASS $H$, and 2MASS $K_s$ catalog fluxes; dashed lines mark equality.}
\label{fig:pointsource_validation}
\end{figure*}

Figure~\ref{fig:pointsource_validation} gives independent point-source validation.  Clean WISE W1, 2MASS $H$, and 2MASS $K_s$ sources have log-space correlations of $r=0.93$, 0.96, and 0.98, with median SPHEREx/reference ratios of 0.98, 0.96, and 0.94.  We do not use diffuse WISE W1 as an absolute reference because bandpass/color terms, zero levels, masking, resolution matching, and WISE processing become entangled; point sources instead check wavelength response, astrometry, and continuum scale.  The diffuse continuum-subtracted feature maps are validated by the feature-specific components listed in Table~\ref{tab:validation_summary}.

\begin{figure*}[!t]
\centering
\includegraphics[width=0.98\textwidth]{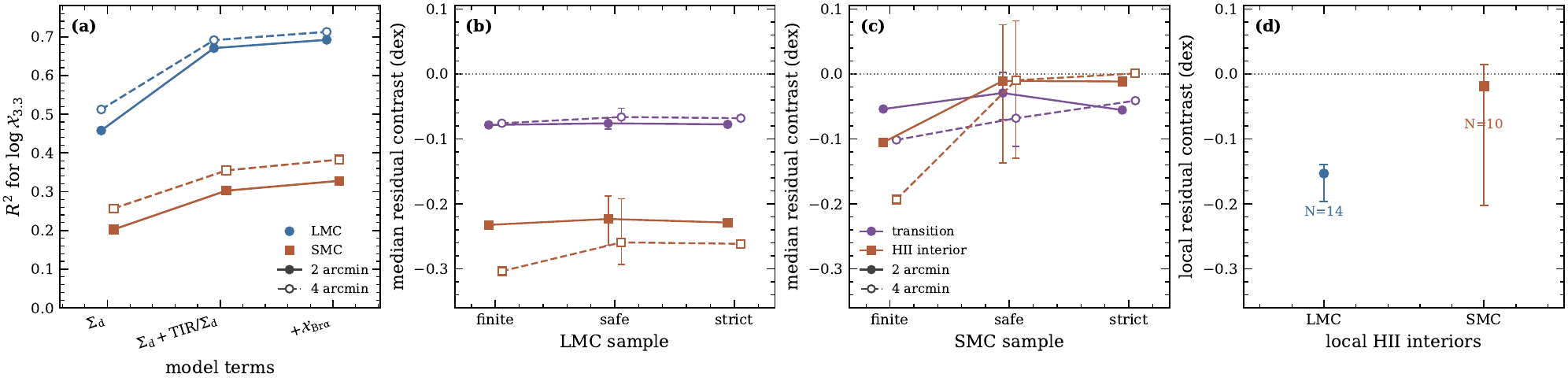}
\caption{Robustness summary for the physical decomposition.  Panel (a) shows nested-model predictive power for \Xthree.  Panels (b) and (c) show median residual contrasts relative to the quiescent diffuse baseline, with 16--84 percentile spatial block-bootstrap intervals for the safe samples.  Panel (d) shows the local-background H{\sc ii}-interior measurement.}
\label{fig:robustness_summary}
\end{figure*}

Figure~\ref{fig:robustness_summary} summarizes physical-decomposition robustness.  Adding heating to dust column strongly improves the \Xthree\ prediction, while adding Br$\alpha$ gives only a small gain.  LMC transition cells and H{\sc ii} interiors remain negative for finite-coverage, fiducial, and strict-coverage samples at 2 and 4 arcmin; the fiducial H{\sc ii}-interior interval is $-0.264$ to $-0.188$ dex at 2 arcmin and $-0.293$ to $-0.192$ dex at 4 arcmin.  SMC intervals are broader and centered near zero.  Heating-proxy variants give the same qualitative LMC result for TIR/$\sigmad$ using BEMBB ($-0.223$ dex) or THEMIS ($-0.182$ dex) dust columns, and for 24 or 160 \um\ intensity per dust mass ($-0.122$ and $-0.186$ dex); 70 \um\ per dust mass and dust temperature alone are weaker proxies.

\begin{figure*}[!t]
\centering
\includegraphics[width=0.65\textwidth]{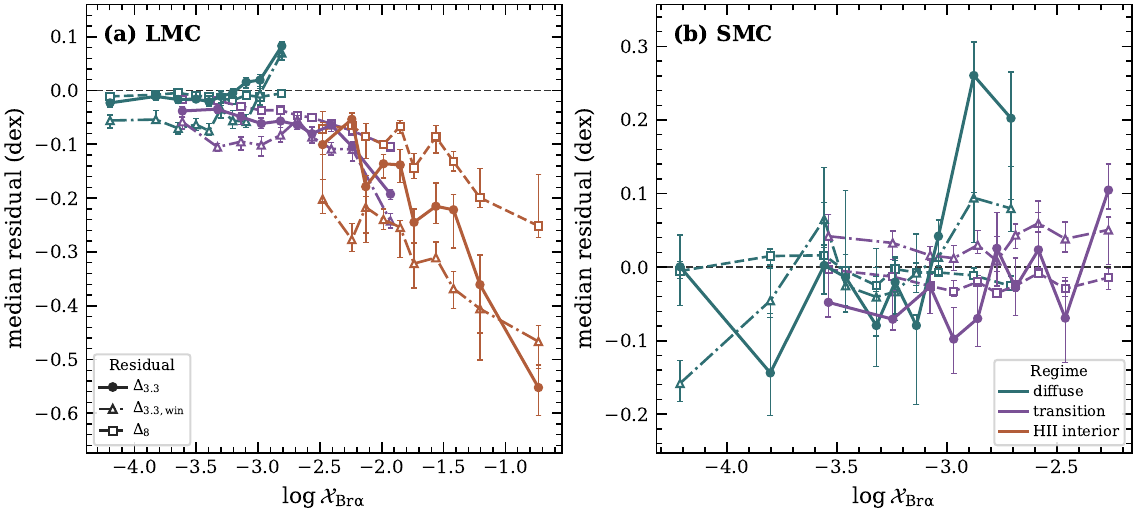}
\caption{Definition-control test of binned median residuals versus \Xbralpha\ for \Xthree, \Ithreewin, and \Ieight.  Symbols and lines identify observables; colors identify environments; vertical bars give 16--84 percentile bootstrap uncertainties.}
\label{fig:i33win_definition_check}
\end{figure*}

Figure~\ref{fig:i33win_definition_check} shows that the LMC H{\sc ii}-interior residual contrasts are $-0.223$ dex for \Xthree, $-0.296$ dex for \Ithreewin, and $-0.102$ dex for \Ieight.  The stronger 3.3 \um\ deficit is therefore not caused by comparing an isolated excess with an 8 \um\ broadband intensity.

\begin{figure*}[!t]
\centering
\includegraphics[width=0.7\textwidth]{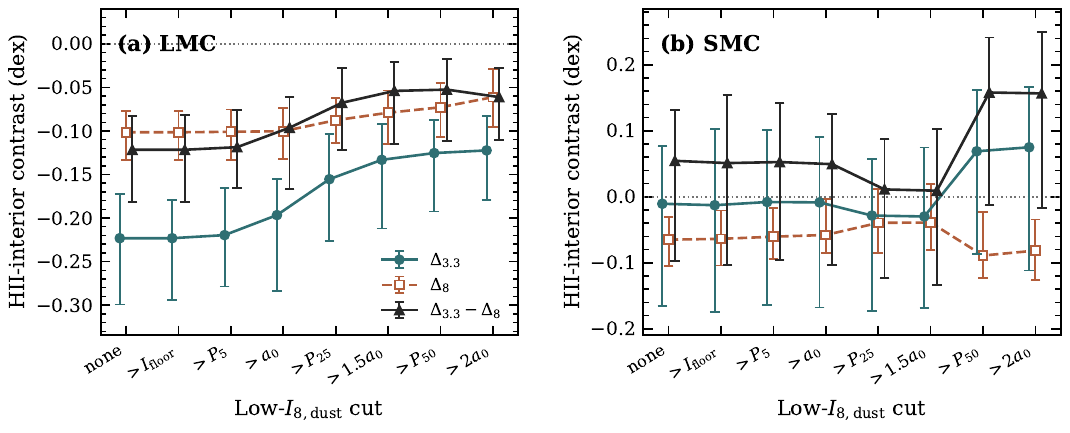}
\caption{Robustness of the matched \Xthree--\Ieight\ residual comparison to low-\Ieight\ cuts.  For each cut, both baselines are refit on the same surviving cells.  Panels show H{\sc ii}-interior contrasts for \Deltathree, \Deltaeight, and their difference; error bars give 16--84 percentile spatial block-bootstrap intervals.  The LMC differential contrast remains negative under all cuts, while SMC points sample lower-\Xbralpha\ cells and remain weak within broader intervals.}
\label{fig:i8_floor_cut}
\end{figure*}

Figure~\ref{fig:i8_floor_cut} tests the apparent low-\Ieight\ floor by refitting both baselines after increasingly strict \Ieight\ cuts.  In the LMC, the H{\sc ii}-interior $\Deltathree-\Deltaeight$ remains negative for every cut, including cuts above the \Ieight\ median or twice the fitted additive-intercept estimate.  In the SMC, the same cuts preserve the lower-\Xbralpha\ comparison and leave the differential residual weak within broader intervals.

\bibliography{references}
\bibliographystyle{aasjournalv7.1}

\end{document}